\RequirePackage{fix-cm}
\documentclass[smallextended]{svjour3}       
\smartqed  
\usepackage{graphicx,amsmath}
\usepackage{nicefrac,color,isotope}
\newcommand{\bea}{\begin{eqnarray}}
\newcommand{\eea}{\end{eqnarray}}
\newcommand{\be}{\begin{equation}}
\newcommand{\ee}{\end{equation}}

 \journalname{Jour. of Low Temp. Phys.}
\begin{document}

\title{Composition of nuclear matter with light clusters and  Bose-Einstein
condensation of  $\alpha$ particles
}

\titlerunning{Light clusters and  Bose-Einstein condensation of
  $\alpha$ particles}        

\author{Xin-Hui Wu         \and
        Si-Bo Wang  \and \\ 
       Armen Sedrakian\and
        Gerd R\"opke 
}


\institute{ Xin-Hui Wu         \and
        Si-Bo Wang  
\at
School of Physics, Peking University, Beijing, 100871, China\\
School of Physics, Nankai University, Tianjin, 300071, China\\
Institute for Theoretical Physics, Goethe-University, D-60438 Frankfurt-Main, Germany\\
              \email{wuxinhui@pku.edu.cn, wsb2016phy@pku.edu.cn}          
           \and
 Armen Sedrakian \at
Frankfurt Institute for Advanced Studies, D-60438 
  Frankfurt-Main, Germany\\
Institute for Theoretical Physics, Goethe-University, D-60438
Frankfurt-Main, Germany\\
\email{sedrakian@fias.uni-frankfurt.de} 
     \and
 Gerd R\"opke\at
Institut f\"ur Physik, Universit\"at Rostock, 
D-18059 Rostock, Germany\\
National Research Nuclear University (MEPhI), 115409 Moscow, Russia\\
\email{gerd.roepke@uni-rostock.de} 
}

\date{Received: date / Accepted: date}
\def\aapr{Astronomy and Astrophysics Reviews}
\def\aj{AJ}%
\def\actaa{Acta Astron.}%
\def\araa{ARA\&A}%
\def\apj{ApJ \ }%
\def\apjl{ApJ \ }%
\def\apjs{ApJS}%
\def\ao{Appl. Opt.}%
\def\apss{Ap\&SS}%
\def\aap{A\&A}%
\def\aapr{A\&A Rev.}%
\def\aaps{A\&A}%
\def\azh{AZh}%
\def\baas{BAAS}%
\def\bac{Bull. astr. Inst. Czechosl.}%
\def\caa{Chinese Astron. Astrophys.}%
\def\cjaa{Chinese J. Astron. Astrophys.}%
\def\icarus{Icarus}%
\def\jcap{J. Cosmology Astropart. Phys.}%
\def\jrasc{JRASC}%
\def\mnras{MNRAS}%
\def\memras{MmRAS}%
\def\na{New A}%
\def\nar{New A Rev.}%
\def\pasa{PASA}%
\def\pra{Phys. Rev. A\ }%
\def\prb{Phys. Rev. B \ }%
\def\prc{Phys. Rev. C\ }%
\def\prd{Phys. Rev. D \ }%
\def\pre{Phys. Rev. E \ }%
\def\prl{Phys. Rev. Lett. \ }%
\def\pasp{PASP}%
\def\pasj{PASJ}%
\def\qjras{QJRAS}%
\def\rmxaa{Rev. Mexicana Astron. Astrofis.}%
\def\skytel{S\&T}%
\def\solphys{Sol. Phys.}%
\def\sovast{Soviet Ast.}%
\def\ssr{Space Sci. Rev.}%
\def\zap{ZAp}%
\def\nat{Nature}%
\def\iaucirc{IAU Circ.}%
\def\aplett{Astrophys. Lett.}%
\def\apspr{Astrophys. Space Phys. Res.}%
\def\bain{Bull. Astron. Inst. Netherlands}%
\def\fcp{Fund. Cosmic Phys.}%
\def\gca{Geochim. Cosmochim. Acta}%
\def\grl{Geophys. Res. Lett.}%
\def\jcp{J. Chem. Phys.}%
\def\jgr{J. Geophys. Res.}%
\def\jqsrt{J. Quant. Spec. Radiat. Transf.}%
\def\memsai{Mem. Soc. Astron. Italiana}%
\def\nphysa{Nucl. Phys. A}%
\def\physrep{Phys. Rep. \ }%
\def\physscr{Phys. Scr}%
\def\planss{Planet. Space Sci.}%
\def\procspie{Proc. SPIE}%

\maketitle

\begin{abstract}
  The Bose-Einstein condensation of $\alpha$ partciles in the
  multicomponent environment of dilute, warm nuclear matter is
  studied. We consider the cases of matter composed of light clusters
  with mass numbers $A\leq 4$ and matter that in addition these
  clusters contains $\isotope[56]{Fe}$ nuclei. We apply the
  quasiparticle gas model which treats clusters as bound states with
  infinite life-time and binding energies independent of temperature
  and density.  We show that the $\alpha$ particles can form a
  condensate at low temperature $T\le 2$ MeV in such matter in the
  first case.  When the $\isotope[56]{Fe}$ nucleus is added to the
  composition the cluster abundances are strongly modified at low
  temperatures, with an important implication that the $\alpha$
  condensation at these temperatures is suppressed.  
\end{abstract}

\section{Introduction}
\label{sec:introduction}

The physics of Bose-Einstein condensation of $\alpha$ particles and
their superfluidity is one of the exciting ``condensed
matter'' aspects of nuclear physics.  The clustering and condensation
in dilute nuclear matter is of interest in nuclear structure
calculations, heavy ion collisions in laboratory experiments as well
as in astrophysics of supernovas and neutron stars. For example, the
processes of neutrino emission and absorption, which are an important
part of the mechanism of core-collapse supernovas and nucleosynthesis,
strongly depend on the composition of warm low-density nuclear
matter~\cite{1990RvMP...62..801B,1996NuPhA.606...95B,2007PhR...442...38J,2008JPhG...35a4056L,2007PhR...442...23B}.


The purpose of this work is to investigate the properties of light-nuclei
in nuclear matter in thermodynamic equilibrium in the density regime
$n\le 0.3\ n_{\rm sat}$, where $n_{\rm sat}\simeq 0.16$ fm$^{-3}$ is
the saturation density, and the temperature regime in the range
$2 \le T\le 10$ MeV. We consider matter composed of a mixture of
clusters and free nucleons. As we are interested in physical processes
which occur on time-scales shorter than the relaxation time needed to
establish $\beta$-equilibrium, we will specify the isospin asymmetry
in terms of proton fraction $Y_p=N_p/(N_p+N_n)$, where $N_p$ and $N_n$
are the net number densities of protons and neutrons.


The determination of the composition and properties of dilute nuclear
matter is a long-standing problem which has gained renewed interest in
recent years in various contexts, for a recent review
see~\cite{2017RvMP...89a5007O}.  The topics of current interest are,
for example, the improvements on the supernova and warm neutron star
equations of state and
thermodynamics~\cite{1991NuPhA.535..331L,1998NuPhA.637..435S} which
include the multi-cluster composition of
matter~\cite{2008PhRvC..77e5804S,2009PhRvC..80a5805H,2009ApJ...707.1495S,2010NuPhA.843...98B,2010PhRvC..81a5803T,2011PhRvC..84e5804H,2012PhRvC..85e5806O,2012ApJ...748...70H,2014ApJ...789...33B,2015PhRvC..92e5804B,2015PhRvC..91e5801P,2016EPJA...52...50O,2017PhRvC..95c5802F}. Another
aspect of the problem is the effects of light clusters in intermediate
energy heavy ion
collisions~\cite{2008NuPhA.809...30D,2009PhRvC..79c4604S,2009PhRvC..79a4002R,2010PhRvL.104t2501N,2014EPJA...50...39H,2017MPLA...3250010M}
which were extensively studied using various methods, see for example
\cite{1991PhLB..260..271P,1994RPPh...57..533P,2016EPJA...52..244B}.
Furthermore, the general many-body problem of bound state formation in
nuclear medium is an outstanding problem on its own
right~\cite{1996PhLB..376....7B,1998AnPhy.266..524S,2006PhRvC..73c5803S,2006NuPhA.776...55H,2012PhRvC..85e5811F,2012PhRvC..86f2801S,2014PhRvC..90f5804S,2014JPhCS.496a2008S,2013NuPhA.897...70R,2015PhRvC..92e4001R,2015PhRvC..92e5803G,2016JPhCS.702a2012C}.

In this work we adopt the quasiparticle gas
model~\cite{2009PhRvC..80a5805H} to explore the composition and
thermodynamics of isospin symmetrical and asymmetrical nuclear matter. We
update and improve on the results of Ref.~\cite{2009PhRvC..80a5805H}
and provide additional information which facilitates a comparison with
the results obtained within alternative models.  The quasiparticle gas
model includes mean-field effects on the nucleon masses (including
those nucleons that are bound in clusters), but neglects the
interactions among clusters with $ A>1$. Thus, the clusters are
treated as quasiparticles with infinite life-time with binding
energies that are independent of the temperature and density.

The main focus of this work is the Bose-Einstein condensation
(hereafter BEC) of $\alpha$ particles in the clustered
environment. The BEC in $\alpha$ matter has attracted much (and
renewed) attention in relation to the problem of $\alpha$ cluster
structure of a number of nuclei, notably
$\isotope[12]{C}$~\cite{2001PhRvL..87s2501T,2002PhRvC..65f4318S}, as
well as $\alpha$ condensation in infinite nuclear
systems~\cite{1966AnPhy..40..127C,1970NuPhA.155..561M,1998PhRvL..80.3177R,2006MPLA...21.2513R,2007PhRvL..98r2501L,2006NuPhA.766...97S,2007PrPNP..59..285S,2008PhRvC..77f4312F,2009PhLB..682...33C,2010PhRvC..82c4322S,2012LNP...848..229Y}. It
is our aim to study this phenomenon in infinite nuclear system under
the condition of chemical equilibrium between $\alpha$ particles and
other light clusters.  We further explore the effect of a heavy
nucleus on the system of light cluster and show that the number
density of $\alpha$ particles reduces in this case and, as a consequence, their
condensation is suppressed.


The paper is organized as follows.  Section~\ref{sec:formalism}
reviews the formalism of quasiparticle gas model adopted in this work
and details the approximations used. In Sec.~\ref{sec:results} we
present the numerical results for the composition and equation of
state of matter. The problem of $\alpha$ particle condensation and the
effect of a heavy nucleus $\isotope[56]{Fe}$ on the cluster abundances
is discussed. A summary and outlook is given in
Sec.~\ref{sec:summary}.


\section{Formalism}
\label{sec:formalism}

\subsection{The quasiparticle gas model}

In this work we consider nuclear matter composed of unbound nucleons
and light nuclei with mass numbers $A\le 4$ plus a single heavy
nucleus. Such an approach was introduced already by
Refs.~\cite{1991NuPhA.535..331L,1998NuPhA.637..435S} and has been used
subsequently in many studies.  The matter is in equilibrium at
temperature $T$ with total number density of nucleons $n$. The nuclear
clusters (bound states) are uniquely characterized by their mass
number $A$ and charge $Z$, which we combine in a single index
$\alpha=(A,Z)$ (no confusion with the $\alpha$ particle should arise).
In the quasiparticle gas model, which assumes that clusters are
particles with infinite life-time, the grand canonical potential
of the ensemble is expanded into partial contributions from individual
constituents (nucleons and clusters) according to
\be\label{eq:thermopotential} 
\Omega (\mu_n,\mu_p, T)
=\sum_{\alpha}\Omega_{\alpha} (\mu_{\alpha}, T), 
\ee 
where $\mu_n$ and $\mu_p$ are the chemical potentials of neutrons and
protons and $\mu_{\alpha}$ is the chemical potential of the cluster
$\alpha$. Baryon number and charge conservation implies that in
chemical equilibrium
\be\label{eq:cheq} \mu_{\alpha} = (A-Z)\mu_n + Z
\mu_p.  
\ee 
The thermodynamic potential for cluster $\alpha$ appearing in 
Eq.~\eqref{eq:thermopotential} is written as 
\be \Omega_{\alpha} (\mu_{\alpha}, T)= -
V\int_{-\infty}^{\mu_{\alpha}} d\mu'_{\alpha}\,\,
n_{\alpha}(\mu'_{\alpha}, T), 
\ee
where $n_{\alpha}(\mu'_{\alpha}, T)$ is the number density of
clusters with a given value of $\alpha$. Here we assume that the
qusiparticle self-energies (which in the quasi-particle picture
are approximated by an effective mass) weakly depend on the density, therefore the
corresponding derivatives can be neglected to leading order.  The
densities of the species are in turn given by
\bea\label{eq:density} 
n_{\alpha} &=& g_{\alpha}\int\frac{ d^3p}{(2\pi)^3} f_{\alpha}(p), 
\eea 
where $g_\alpha=2s_\alpha+1$ is the degeneracy factor for the spin degree
of freedom and  $f_{\alpha}(p)$ are the Fermi/Bose distribution functions
\bea
f_\alpha(p)=\left[\exp\left(\frac{E_\alpha-\mu_\alpha}{T}\right)
-(-1)^{A}\right]^{-1},\eea
where the quasiparticle energies are given by 
\bea
E_\alpha= \left\{ 
\begin{array}{ll}
\frac{p^2}{2m^*} ,& \textrm{unbound nucleons}\\
\frac{p^2}{2m^*A}-B_\alpha, & \textrm{clusters}
\end{array}\right.
\eea
where $p$ is the momentum of a cluster in its center-of-mass frame,
$m^*$ is the nucleon effective mass and 
$B_\alpha$ is the binding energy. Here and in the following we neglect
the small mass difference between proton and neutron.  

The key feature of the quasiparticle gas model is that the density is
a sum of contributions from infinite life-time quasiparticles -
clusters - characterized by the value of index $\alpha$.  The
thermodynamic quantities of interest now can be computed from the
thermodynamic potential Eq.~(\ref{eq:thermopotential}). The pressure
and the entropy are obtained as
\begin{equation}
P= -\frac{\Omega}{V} , \quad \quad  
S = -\frac{\partial \Omega}{\partial T}.
\end{equation}
It is clear that the pressure of the system is the sum over pressures
of each type of clusters present in the mixture: $P=\sum_\alpha
P_\alpha$, where 
\bea\label{eq:pressure}
  P_\alpha =-\frac{\Omega_{\alpha}}{V}=\pm \frac{g_\alpha T }{2\pi^2}
\int^\infty_0 \ln\left[1\pm
  \exp\left(\frac{\mu_\alpha-E_\alpha}{T}\right)\right]
\ k^2  \ dk. 
\eea
Similarly, the net entropy of the system is given by 
$S=V^{-1}\sum_\alpha S_\alpha$, where 
\bea
\label{eq:entropy} 
S_{\alpha}=\mp \frac{g_\alpha
  V}{2\pi^2}\int^\infty_0 [f_{\alpha}\ln f_{\alpha}
+(1\mp f_{\alpha})\ln (1\mp f_{\alpha})] \ k^2 \ dk.
\eea
Finally, we establish the relations between the number densities of 
clusters given by Eq.~\eqref{eq:density} and  the total proton
number density $N_p$, the total neutron number density $N_n$
at temperature $T$
\bea 
    N_n(T,\mu_p,\mu_n)&=&\sum_\alpha (A-Z)n_\alpha(T,\mu_p,\mu_n),\\
    N_p(T,\mu_p,\mu_n)&=&\sum_\alpha Zn_\alpha(T,\mu_p,\mu_n) . 
\eea 
We will use below as independent variables the total nucleon number density
and isospin asymmetry parameter defined as 
\be 
  n=N_n+N_p,\quad Y_p=N_p/n.
\ee

\subsection{Bose-Einstein condensation of $\alpha$ particles}
\label{sec:BEC}

The formalism described above fully accounts for the quantum
statistics of the clusters, therefore any putative BEC in clustered
matter is automatically included.  The expression for the densities of
clusters Eq.~\eqref{eq:density} is not valid in the case of BEC,
because in this case macroscopic number of particles occupy the ground
state with zero momentum.  If the number of particles occupying the
zero-momentum mode is $n_0$ then cluster density is given by
\bea
  n_\alpha = n^0_{\alpha}
+\frac{g_\alpha }{2\pi^2}\int_{0^+}^\infty f_\alpha k^2 dk,
\eea
where $0^+$ indicates exclusion of the zero-momentum mode from the
intergal and the index $\alpha$ may refer to the $\alpha$ particle as well
as to the deuteron $d$.  The chemical potential of bosonic clusters lies in
the interval $-\infty \le \mu_\alpha\le -B_{\alpha}$. The condition of
the Bose condensation is achieved at the upper limit, where the
chemical potential approaches the binding energy of the cluster. The
temperature corresponding to the limit $\mu_{\alpha } = - B_{\alpha}$
for fixed density is identified as the critical temperature $T_c$  of
BEC.  This temperature can be determined
from the transcendental equation
\bea n_{\alpha}=\frac{g_\alpha }{2\pi^2}\int_0^\infty\!\!
\frac{k^2\ dk }{\exp\left(\frac{k^2}{2Am^*T}\right)-1}\eea
which leads to the critical temperature 
\bea\label{tc} 
T_{c\alpha} = \frac{2\pi}{Am^*}\left(\frac{n_{\alpha}}{g_\alpha\zeta(3/2)}\right)^{2/3},
\eea
where the Riemann zeta-function has the value $\zeta (3/2) = 2.612 $,
$A=4$ and $g_\alpha=1$  for the $\alpha$ particle.

With the onset of the BEC the number density of the condensed clusters
does not depend on the chemical potential, as it is exactly canceled
by the contribution from the binding energy.  On the other hand the
conversion from one type of a cluster  to another is controlled through
the chemical equilibrium condition \eqref{eq:cheq}.  Thus, after
clusters of type $\alpha$ condense the equation determining their
number density is decoupled from the rest of the system and
inter-cluster transformations from and to the condensate will not
occur. This reflects the fact that after the onset of BEC the
condensate particles in the ground state do not interact with the
environment. Therefore, at a given temperature, an increase in the density
of the system will result only in an increase of the density of the
condensate, whereas the densities of the remaining clusters will be
frozen.

\subsection{Further approximations}

The effective mass of nucleons is determined from 
the Skyrme functional and is given by 
\bea 
\frac{m^*}{m} &=&\Big\{1+\frac{mn}{2}(t_1+t_2)+\frac{mn}{8}(t_2-t_1)[1\pm(1-2Y_p)]\Big\}^{-1},
\eea 
where the parameters $t_0,t_1$ and $t_2$  have the following values
$t_0=-1128.75$ $\mathrm{MeV~fm^3}$,
$t_1=395$ $\mathrm{MeV~fm^5}$, and 
$t_2=-95$ $\mathrm{MeV~fm^5}$.
Note that the effective mass of a nucleon is used uniformly both for
unbound nucleons and clusters, but its actual value is very close to
unity at low densities of interest. We neglect the weak density
dependence of the effective mass in evaluating the thermodynamical
potential of the system.  In the following we will also neglect the
medium modifications of the binding energies of clusters, i.e., their
dependence on the temperature and density of the ambient matter, which
is justified at densities below $n_{\rm sat}/3$. These modifications
are discussed
elsewhere~\cite{2010PhRvC..81a5803T,2011PhRvC..84e5804H,2009PhRvC..79a4002R,2006PhRvC..73c5803S,2015PhRvC..92e4001R}. The
numerical values of the binding energies of light clusters used in our
computations are $B_d=2.225$ (deuteron), $B_t=8.482$ (triton),
$B_h=7.718$ (helion) and $B_{\alpha}$ =28.3 MeV ($\alpha$ particle).
The degeneracy factors are $g_n=g_p=g_t=g_h=2$, $g_d=3$ and
$g_\alpha=1$.

\begin{figure}[t] 
\begin{center}
\includegraphics[width=0.75\textwidth,keepaspectratio]{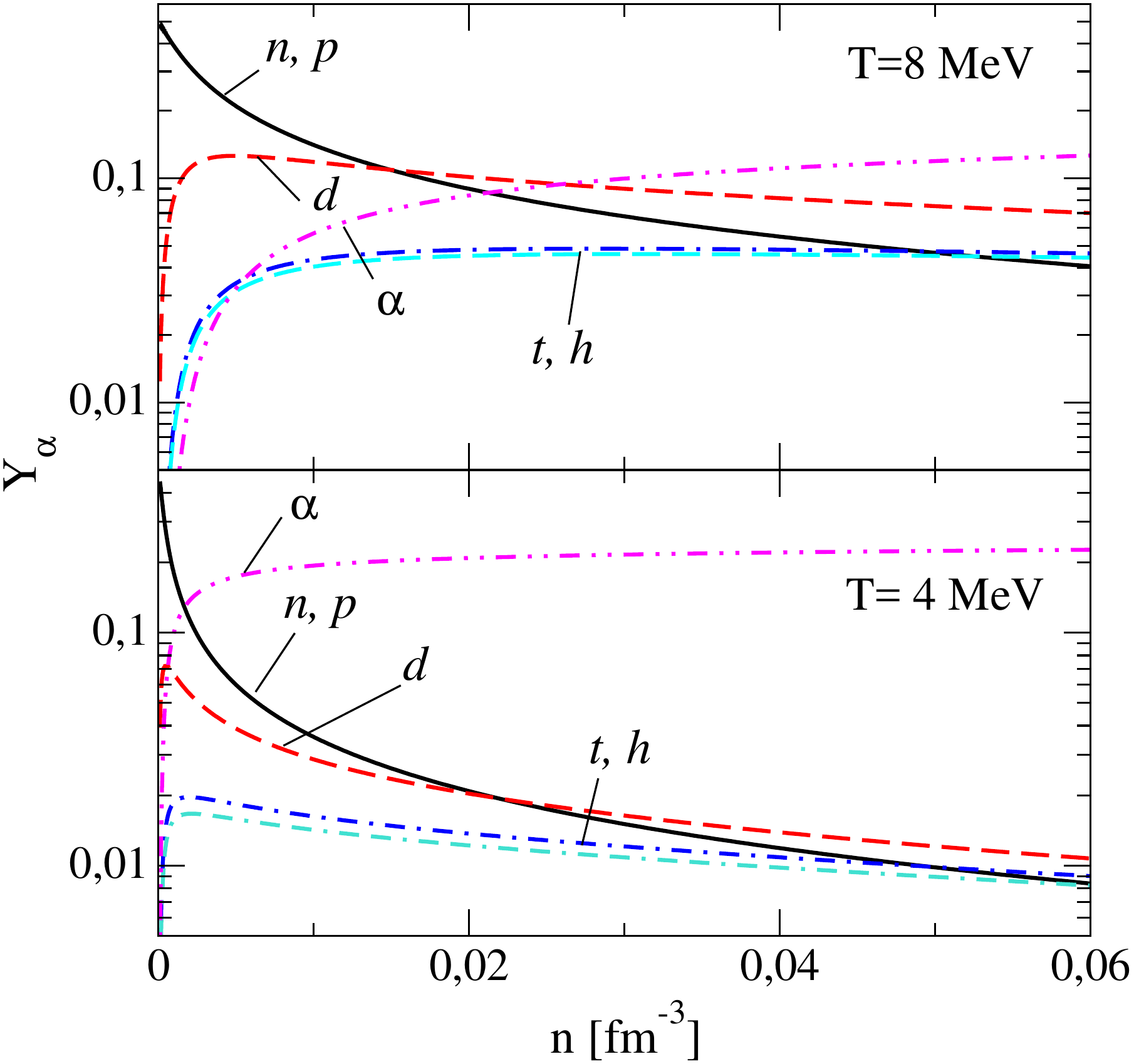}
\caption{ Dependence of abundances of nuclei $Y_{\alpha}=n_{\alpha}/n$ with
  $A\leq 4$ on matter density in isospin symmetrical matter at
  temperatures $T=10~\mathrm{MeV}$ (upper panel) and
  $T=4~\mathrm{MeV}$ (lower panel).  }
\label{fig:y_vs_dens}
\end{center}
\end{figure}
\section{Results}
\label{sec:results}

\subsection{Light clusters only}

The system of coupled non-linear equations \eqref{eq:density} (one
equation per cluster) was solved simultaneously for unknown chemical
potentials~$\mu_n$ and $\mu_p$ at fixed temperature $T$, number
density $n$ and asymmetry parameter $Y_p$. Consider first symmetric
nuclear matter with $Y_p=0.5$ composed of neutrons ($n$), protons
($p$), deuterons ($d$), tritons ($t$), helions ($h$) and
$\alpha$ particles ($\alpha$) and denote their abundance as 
$Y_{\alpha } = n_{\alpha}/n$.

Fig.~\ref{fig:y_vs_dens} shows the abundances of clusters at two
temperatures as a function of the net density.  We recall that in our
model the nuclear interaction renormalizes the nucleon masses
and no further inter-cluster interactions are taken into account. This
should be a good approximation valid at low densities of
interest. Therefore, the system minimizes the energy of a collection
of clusters with various masses, which determine the magnitude of the
their kinetic energy.  Consequently, the competition is between
binding the nucleons in a cluster and moving it with relatively low
velocity (because of the larger mass of the cluster) or keeping
nucleons unbound but moving them with larger velocities.  It is seen
that at high temperatures and low densities the ensemble chooses to
have larger number of nucleons and deuterons, i.e., many constituents
with large velocities.  At lower temperatures $\alpha$ particles
dominate, i.e., smaller number of more massive constituents with lower
velocities is preferable.

\begin{figure}[t]
\begin{center}
\includegraphics[width=0.75\textwidth,keepaspectratio]{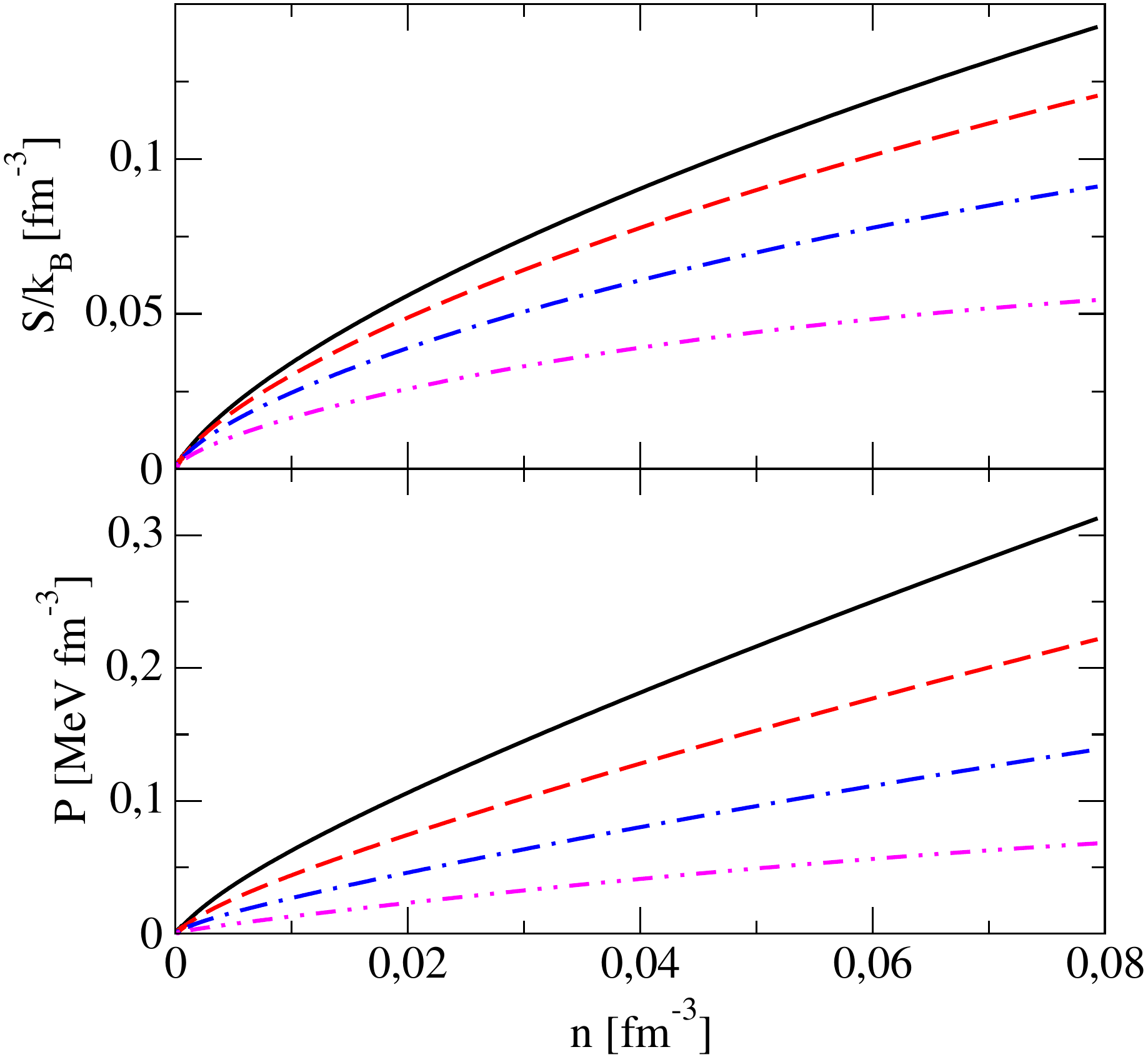}
\caption{ Dependence of entropy (upper panel) and pressure (lower
  panel) of the system on matter density in the temperature range from
  at $T=10$ (upper curve in each panel) MeV to $4$ MeV (lowest curve
  in each panel). The size of the temperature step is 2 MeV.  }
\label{fig:SandP_vs_dens}
\end{center}
\end{figure}

In Fig.~\ref{fig:SandP_vs_dens} we show the entropy and pressure of
our thermodynamic ensemble for various constant temperatures and
varying density. The dependence of these quantities is clearly
consistent with the earlier findings on the thermodynamics of the
quasiparticle gas model~\cite{2009PhRvC..80a5805H}.

\begin{figure}[t] 
\begin{center}
\includegraphics[width=0.75\textwidth,keepaspectratio]{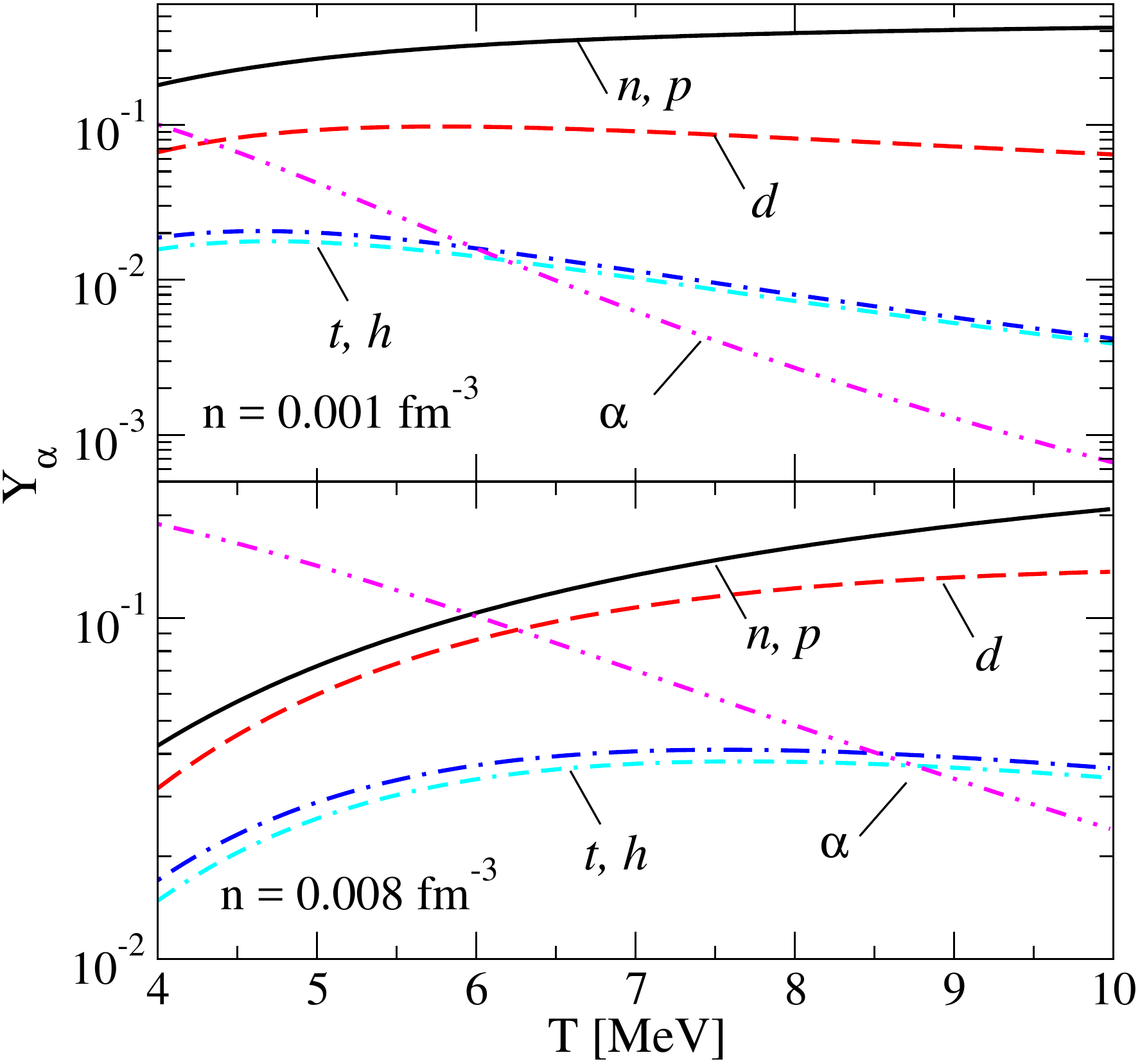}
\caption{ Dependence of components of symmetrical nuclear matter on
  temperature at fixed total density $n=0.001$~fm$^{-3}$ (upper panel)
  and $n=0.008$~fm$^{-3}$ (lower panel).  }
\label{fig:Y_vs_T}
\end{center}
\end{figure}

We turn now to the cluster abundances at fixed density and varying
temperature, as shown in Fig. \ref{fig:Y_vs_T}. For the lower of the
 chosen values of
density the nucleons and deuterons dominate the matter
composition. Among heavier clusters $\alpha$ particles dominate at low
temperatures, whereas $t$ and $h$ are more abundant at high
temperatures.  For the higher value of the density and low temperatures $\alpha$
particles completely dominate the matter, which opens up the
possibility of their condensation at a critical temperature. As the
temperature is increased the general trends seen at low density
remain, however $\alpha$ particles start to dominate matter already
at higher temperature and the three-body clusters overtake the $\alpha$
particle only at the upper edge of the temperature range considered.

\begin{figure}[t] 
\begin{center}
\includegraphics[width=0.72\textwidth,height=7.7cm]{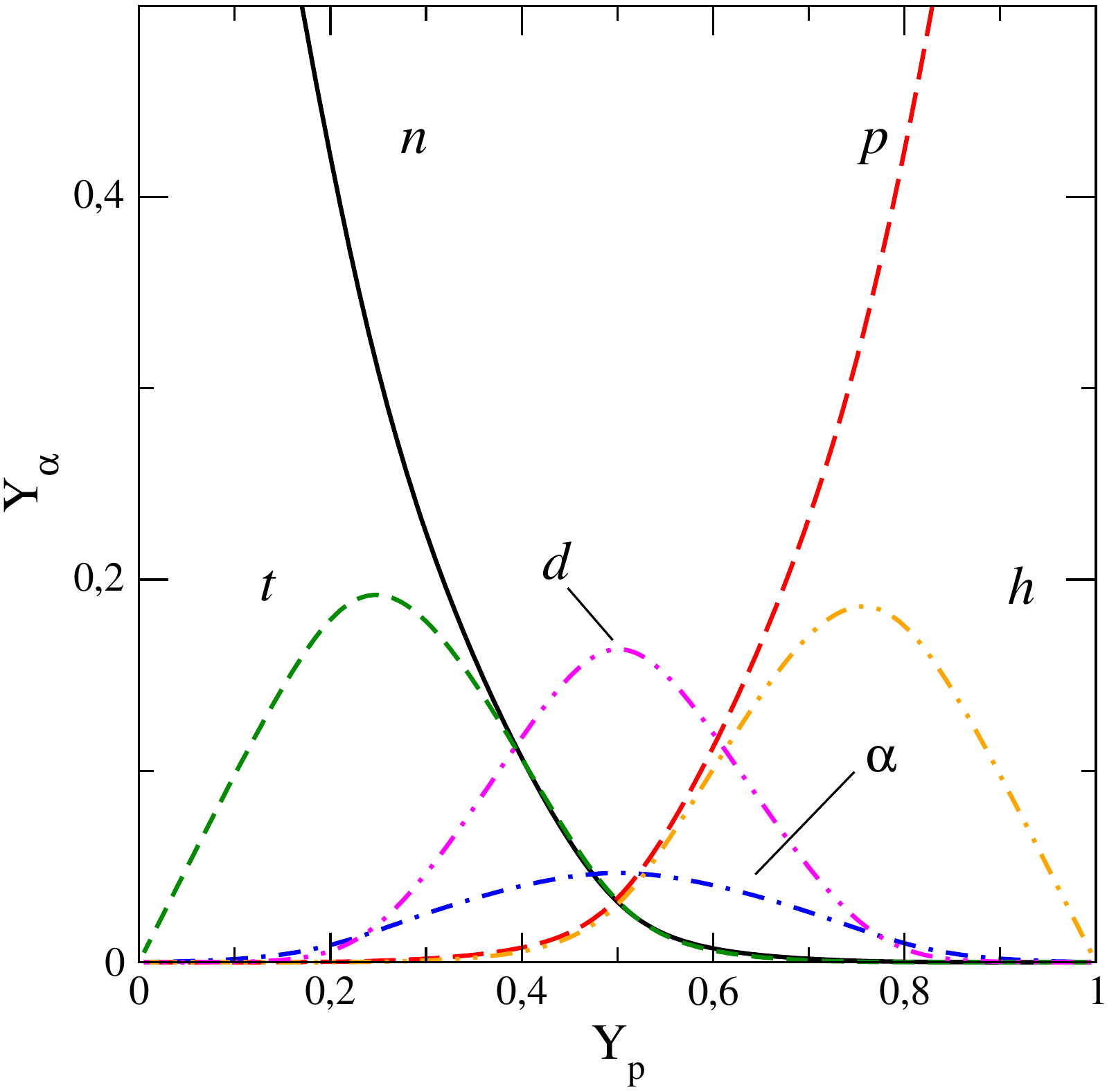}
\caption{ Dependence of composition of asymmetrical nuclear matter on
  parameter $Y_p\in(0,1)$ at fixed total density
  $n=0.04$ fm$^{-3}$ and temperature $T=6$ MeV.  }
\label{fig:Ya_vs_Ye}
\end{center}
\end{figure}
\begin{figure}[!] 
\begin{center}
\includegraphics[width=0.75\textwidth,height=7.8cm]{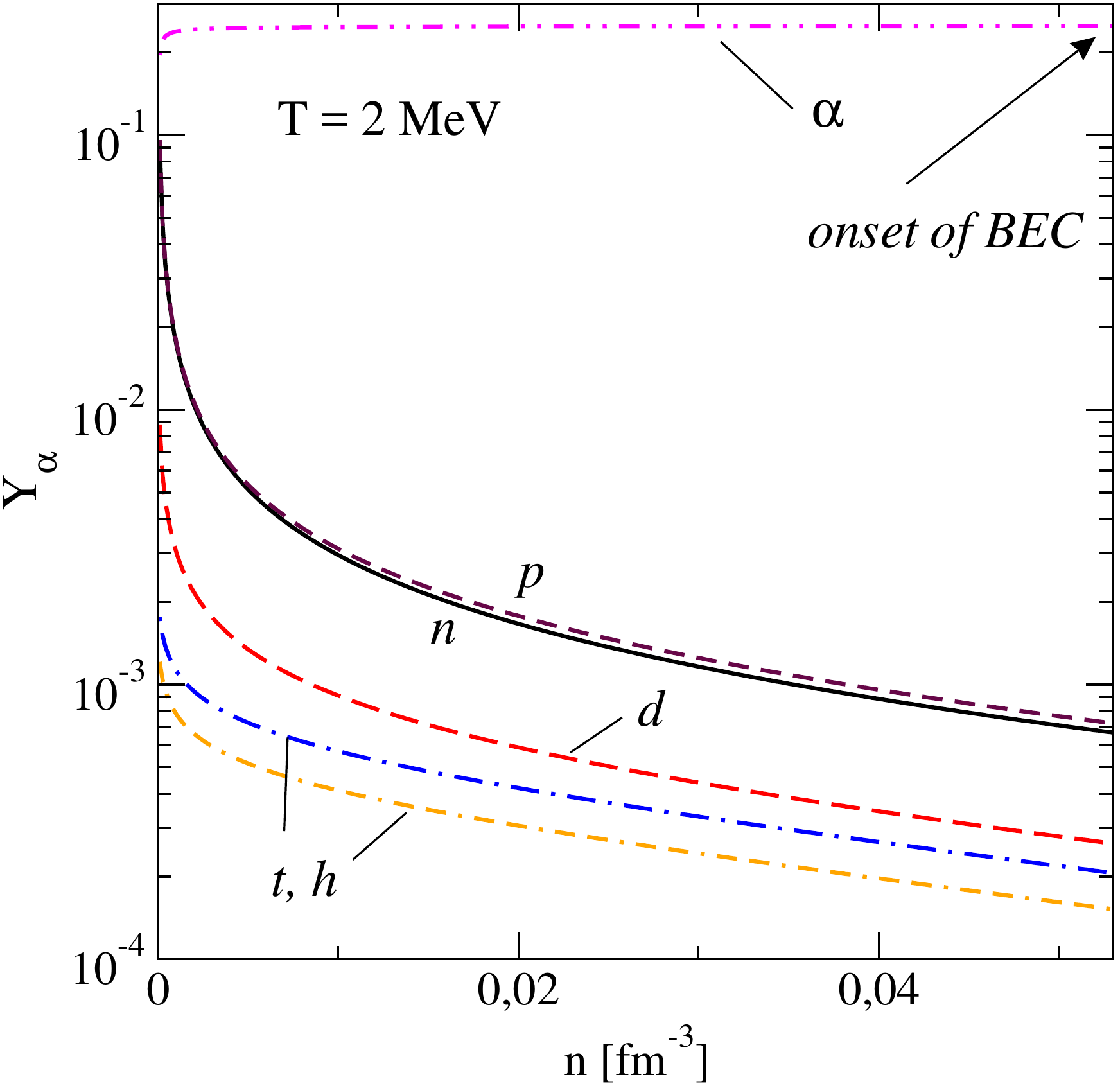}
\caption{ Dependence of abundances of components $(A\leq 4)$ on total
matter density for temperature at $T=2$ MeV. The onset of BEC occurs
at the maximal density shown in this figure. }
\label{fig:Ya_mua_T2}
\end{center}
\end{figure}
The current set-up allows us to explore the dependence of the
composition on the isospin asymmetry, see Fig.~\ref{fig:Ya_vs_Ye}.
The parameter $Y_p$ is varied in the range $0\le Y_p\le 1$, the lower
limit corresponding to pure neutron matter, $Y_p=0.5$ - to symmetrical
nuclear matter and $Y_p =1$ - to pure proton matter.  The deuteron and
$\alpha$ particle have largest abundances in symmetrical nuclear
matter, i.e., $Y_p=0.5$.  The triton and helion show symmetrical
(mirror) behaviour with respect to the $Y_p=0.5$ value (the slight
numerical difference arises from the difference in their binding
energies).  The decrease in $\alpha$ and $d$ abundances away from the
symmetrical limit is the consequence of the asymmetry in nucleon
number which suppresses binding in isospin symmetrical bound
states. In other words, for a given temperature and density the number
of neutrons for $Y_p>0.5$ and the number of protons for $Y_p<0.5$ is
insufficient to built $\alpha$ particles or $d$ in the same amount as
in the symmetrical case. The maxima in the abundances of $t$ and $h$
lie away from the symmetric limit and reflect the fact that on the
neutron rich side it is easier to built a $t$ (with two neutrons and a
proton) and conversely on the proton rich side it is easier to create
an $h$.

\begin{figure}[!] 
\begin{center}
\includegraphics[width=0.75\textwidth,keepaspectratio]{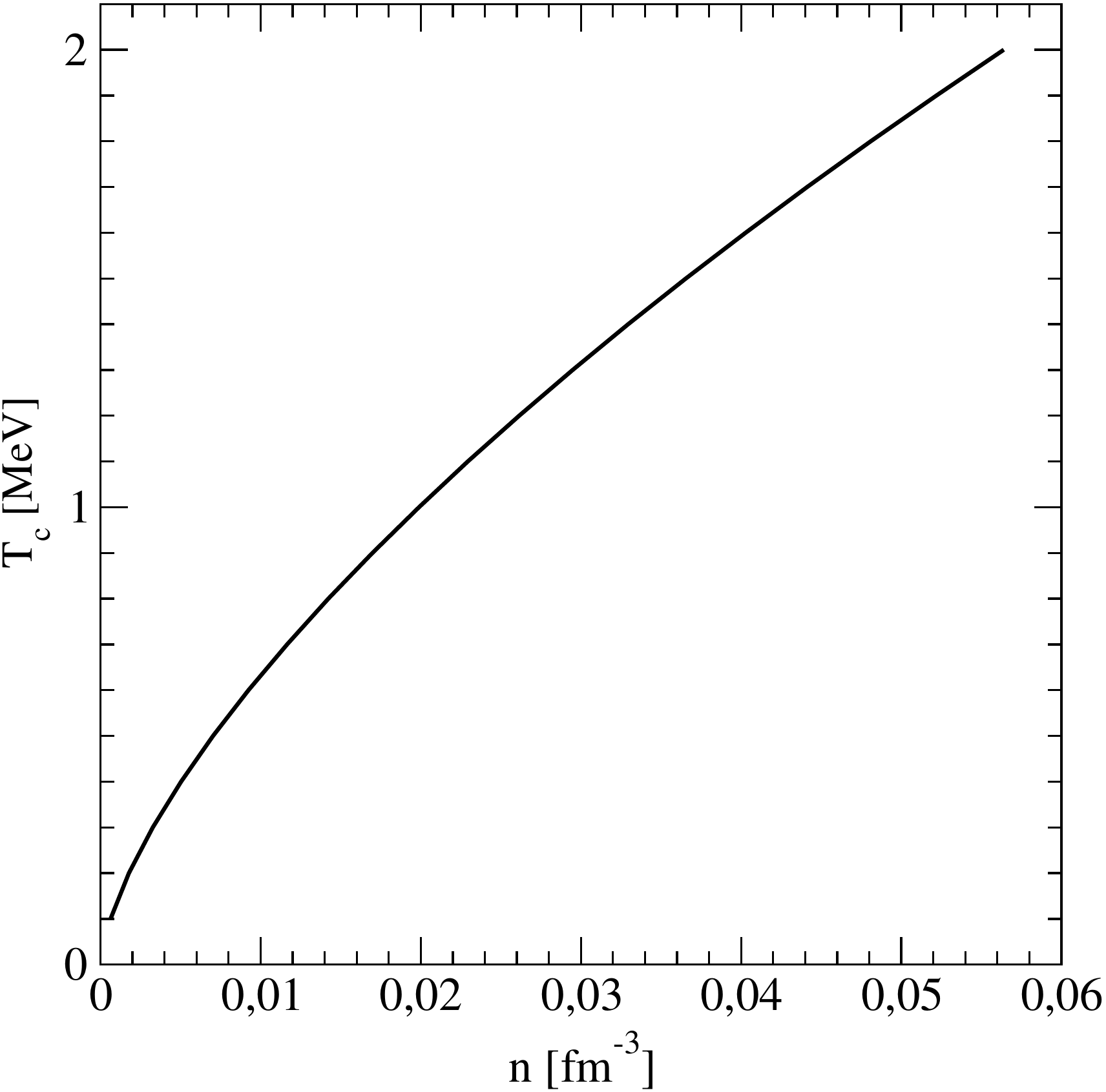}
\caption{Dependence of the critical temperature of condensation of
  $\alpha$ partciles on density. 
}
\label{fig:Tcalpha}
\end{center}
\end{figure}
Having established that at low enough temperatures $\alpha$ particles
form the dominant component of matter we now turn to their BEC.  In
Fig.~\ref{fig:Ya_mua_T2} we show the abundances of clusters up to the
point of the onset of BEC, which is marked by the condition
$\mu_{\alpha}=-B_{\alpha}$.  Once condensed $\alpha$ particles
decouple from the other components and any increase in the number of
nucleons can be accommodated in the condensate (in the isospin
symmetrical matter).  The variation of the critical temperature of BEC
of $\alpha$ particles with density is shown in
Fig.~\ref{fig:Tcalpha}. Clearly, because $\alpha$ particles are not
interacting we are dealing essentially with the condensation in an
ideal quantum Bose gas; as a consequence the condensate fraction in
$\alpha$ matter is controlled by the temperature and all particles
condense in the ground state at $T=0$. The condensate fraction may
reduce substantially if the $\alpha$-$\alpha$ interactions are taken
into account~\cite{1966AnPhy..40..127C,1970NuPhA.155..561M}.

\begin{figure}[t] 
\begin{center}
\includegraphics[width=0.75\textwidth,keepaspectratio]{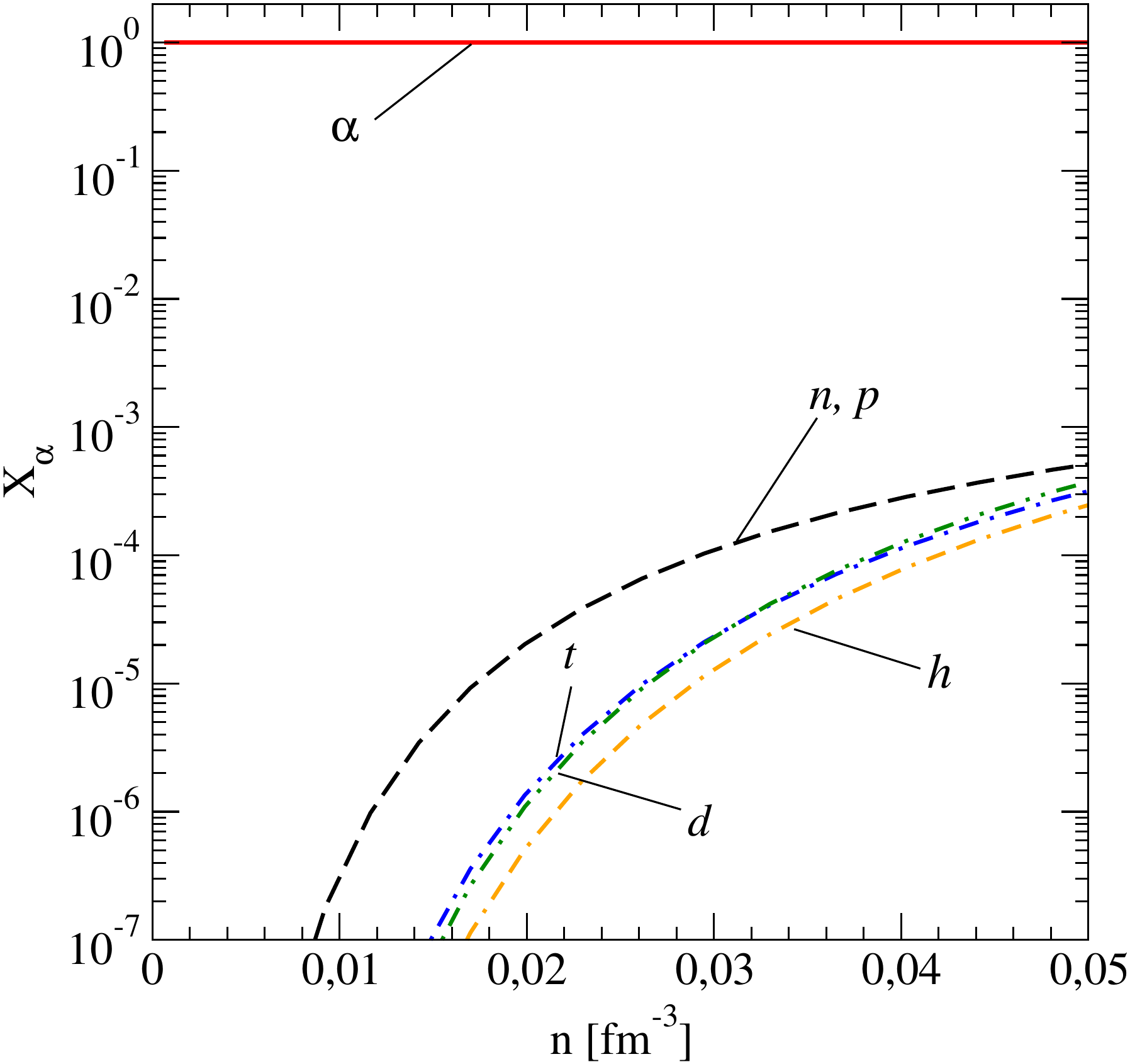}
\caption{ Dependence of the mass fraction $X_{\alpha}$ of light
  clusters on the density at the condensation temperature  for $\alpha$ particles.
}
\label{fig:Xc}
\end{center}
\end{figure}

In Fig. ~\ref{fig:Xc} we show the mass fractions
$X_{\alpha}=An_{\alpha}/n$ of clusters along the trajectory defined
by $T_c(n)$ for $\alpha$ particles.  As can be anticipated from the
discussion of the previous figure we find the mass fractions of
$A\neq 4$ quasiparticles are negligible compared to those of $\alpha$
particles. Even at the highest density their mass fractions do not
exceed $1\%$ of the total mass.

\begin{figure}[t] 
\begin{center}
\includegraphics[width=0.75\textwidth,keepaspectratio]{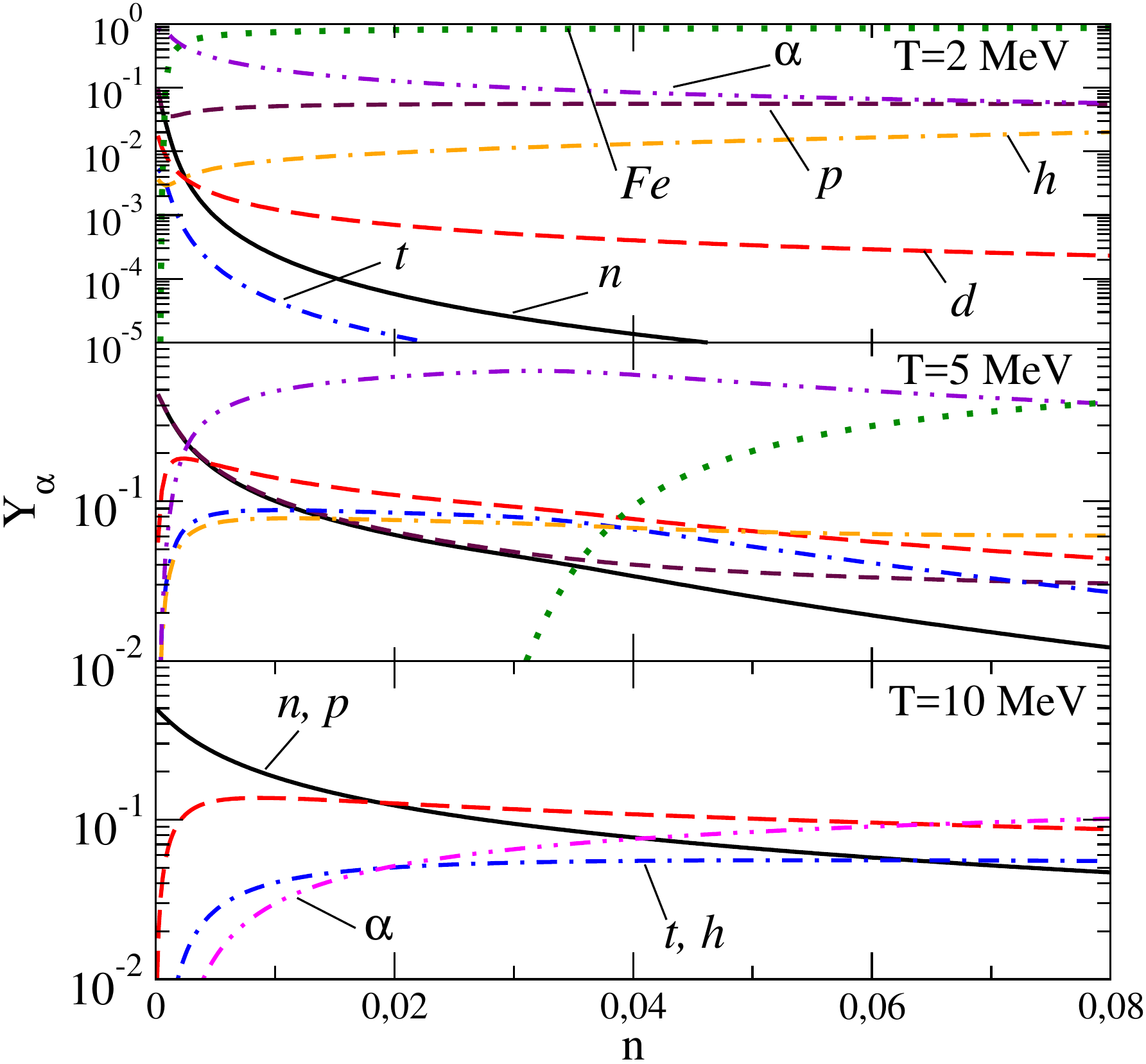}
\caption{Dependence of abundances of light clusters and
 $\isotope[56]{Fe}$ nucleus on density for  $T=2$ MeV (upper panel), 
   5 MeV (middle panel) and 10 MeV (lower panel). 
}
\label{fig:Ya_vs_dens_Fe}
\end{center}
\end{figure}

\subsection{Including a heavy cluster}

As well known, the $\isotope[56]{Fe}$ nucleus is one of the most
tightly bound nuclei (8.8 MeV binding energy per nucleon), therefore
one can anticipate that at asymptotically low densities matter will
consist of iron nuclei.  To access their role on the composition of
matter with light clusters we have added the contribution from
$\isotope[56]{Fe}$ to thermodynamic
potential~\eqref{eq:thermopotential} and recomputed the abundance of
the ensemble consisting of light clusters plus iron.

Figure \ref{fig:Ya_vs_dens_Fe} shows these abundances for several
fixed temperatures as a function of number density. The main
modification arises at low temperatures; it is seen that at $T=2$ the
main element in matter is $\isotope[56]{Fe}$, followed by $\alpha$
particle with about $10\%$ throughout most of the density range. An
exception is the extreme low-density asymptote where the roles of
these elements are interchanged. Interestingly, because
$\isotope[56]{Fe}$ contains 30$n$ and 26$p$ it introduces an isospin
asymmetry even in symmetrical matter and, consequently, the degeneracy
in abundances of $t$ and $h$ as well as $n$ and $p$ is lifted. It is
seen that for a given number of $\isotope[56]{Fe}$ nuclei the $h$ and
$p$ abundances are much larger than those of $t$ and $n$, which is in
agreement with the fact that there is a proton excess.  As the
temperature is increased to 5 MeV the abundance of $\isotope[56]{Fe}$
decreases strongly below densities $0.04$ fm$^{-3}$ and it has no
influences on the low density asymptotics of the abundances, while
still contributing substantially at densities above $0.06$
fm$^{-3}$. Finally, at temperature $T=10$ MeV there is a negligible
amount of $\isotope[56]{Fe}$ nuclei and we recover the composition
studied initially.  The dominance of $\isotope[56]{Fe}$ nuclei in the
low-temperature limit has an important consequence on the $\alpha$
particle condensation: as we have seen it suppresses the number
density of $\alpha$ particles and consequently the conditions for
their condensation are not met anymore. We find that $\alpha$
condensation is completely suppressed if $\isotope[56]{Fe}$ nuclei are
allowed in the composition of matter. Nevertheless, in a number of
situations the time-scales of formation of heavy nuclei such as
$\isotope[56]{Fe}$ may be too long (as, for example, in heavy-ion
collisions or at some stages of supernova explosions) so that the
matter could be composed predominantly of light $A\le 4$ nuclei. In
that case BEC of $\alpha$ particles can indeed take place in clustered
matter as argued above.


\section{Summary and outlook}
\label{sec:summary}

We have calculated the composition of isospin symmetric and asymmetric
warm dense nuclear matter within the framework of quasiparticle gas
model.  The composition of the nuclear matter with light clusters and
unbound nucleons is determined by solving nonlinear equations for
their densities subject to the condition of chemical
equilibrium. Because the quantum-statistic of clusters is taken into
account from the outset we are in a position to monitor possible BEC
of bosons in general.  Our findings can be summarized as follows:
\begin{itemize}
\item At high temperatures $T\simeq 10$ MeV the main component of
  matter are nucleons and deuterons at low densities. At higher
  densities we find a mixture of comparable numbers of $A\le 4$
  clusters. At low temperatures $\alpha$ particles dominate the
  composition of matter above certain density; the lower the
  temperature the lower is the density at which they start to
  dominate.

\item We find BEC of $\alpha$ particles in a clustered environment for
  $n \le 0.3\ n_ {\rm sat} $ with critical temperature $T_c \le 2$ MeV.
  However, in this temperature-density range the abundances of
  the ambient clusters and their effect on the BEC are found to be
  negligible. Therefore, we observe essentially BEC in
  ``alpha-matter''.  As a consequence the critical temperature of
  condensation scales as $T_c \simeq n_\alpha^{2/3}\simeq n^{2/3}$.
  Note that the condition for the occurence of BEC is given by the
  requirement of equality of $\alpha$ particle chemical potential to
  its negative binding energy, as it would be in an non-interacting Bose gas.

\item The addition of a heavy nucleus (in our numerical example
  $\isotope[56]{Fe}$) modifies the composition of matter substantially
  only in the low-temperature limit. This has, however, an important
  consequence of reducing the density of $\alpha$ particles and
  suppressing the $\alpha$ condensation phenomenon. Nevertheless, the
  short time scale dynamics of heavy-ion collisions and supernovas may
  not allow for formation of heavy nuclei, in which case matter
  composed of $A\le 4$ clusters will undergo BEC of $\alpha$
  particles. It may be also of interest under inhomogeneous
  conditions, where heavier nuclei form a separate phase. We
  anticipate that the $\alpha$-$\alpha$ interactions can be neglected
  in a first approximation in the dilute limit relevant for present
  discussion; the influence of interactions on the physics discussed
  above needs further assessment.

\item Note that the medium effects, not considered here, will make the
  formation of a quantum condensate more difficult at high densities
  because Pauli-exclusion principle will act among the nucleons and
  because the $\alpha$-$\alpha$ interactions will reduce the
  condensate fraction in $\alpha$ matter. It should be kept in mind
  that $\alpha$ condensate will likely exist in a transitent form and
  will disappear in thermodynamic equilibrium, as the matter will tend
  to form heavier clusters. Therefore, dynamical (nonequilibrium)
  treatment of the onset of $\alpha$ condensation may be of great
  interest on its own right. 

\item BEC of $\alpha$ particles may be also of interest under
  inhomogeneous conditions, where heavier nuclei form a separate
  phase.

\end{itemize}

The change in the composition of matter as the temperature is lowered
from $T\simeq 10$ MeV to a few MeV may have interesting implications
in astrophysics of compact star mergers and supernovas. The dynamical
properties of matter such as electron transport or neutrino
propagation may be affected by the crossover from matter dominated by
nucleons to matter dominated by $\alpha$ particles, assuming that
heavy nuclei have not been formed. Consequently, the computations of
neutrino opacity~\cite{2016EPJWC.10906002F} or electronic MHD
resistivity~\cite{2016PhRvC..94b5805H} should include such
transformations. The $\alpha$ particle BEC and its influence on the
physical processes in these contexts remain to be studied.

\begin{acknowledgements}
  We thank John Clark, Eckhard Krotscheck, Peter Schuck, Horst
  St\"ocker and Stefan Typel for useful discussions.  X.-H. W. and
  S.-B. W. acknowledge the hospitality of Goethe University, Frankfurt
  am Main, where this project was carried out and Nankai University for
  the support through the International Exchange Foundation.  A. S. is
  supported by the Deutsche Forschungsgemeinschaft (Grant No. SE
  1836/3-2) and by the NewCompStar COST Action MP1304.
\end{acknowledgements}

\bibliographystyle{spphys}

\end{document}